\documentclass{ws-procs9x6}
\usepackage{amsmath,amssymb,amscd,amsfonts,bm}

\newcommand{\la}{\langle}
\newcommand{\ra}{\rangle}

\newcommand{\as}{\alpha_s}
\newcommand{\aspi}{\frac{\alpha_s}{2\pi}}
\newcommand{\La}{\mathrm{a}}
\newcommand{\Lb}{\mathrm{b}}

\def\sket#1{\big|{#1}\big)}
\def\sbra#1{\big({#1}\big|}
\makeatletter
\DeclareRobustCommand*\cal{\@fontswitch\relax\mathcal}

\let\@preprint\relax
\newcommand\preprint[1]{
  \long\gdef\@preprint{\normalsize #1}
} 

\def\@maketitle{%
  \newpage
  \null
  \vspace*{-10pt}
   \vspace*{-25pt}% to have the bylines on the beginning of the page
  \@clinebuf
  \begin{flushright}
  \@preprint
  \end{flushright}
  \vspace*{30pt}
  \begin{center}%
  \let\footnote \thanks
  {\titlefont\@title \par}%
  \vspace*{28pt}
  \@aabuffer\par
  \end{center}%
}
\makeatother

\begin{document}
\preprint{ZH-TH 01/06, hep-ph/0601021}

\title{A new parton shower algorithm: Shower evolution, matching at leading and next-to-leading order level\footnote{\uppercase{C}ontribution to the 
\uppercase{P}roceedings of the \uppercase{R}ingberg workshop on 
``\uppercase{N}ew \uppercase{T}rends in \uppercase{HERA}
\uppercase{P}hysics 2005''. \uppercase{T}his talk was given by 
\uppercase{Z}olt\'an \uppercase{N}agy.}}

\author{Zolt\'an Nagy}

\address{
Institute for Theoretical Physics,
University of Z\"urich \\
Winterthurerstrasse 190, 
CH-8057 Z\"urich,
Switzerland \\ 
E-mail: nagyz@physik.unizh.ch
}

\author{Davision E. Soper}

\address{
Institute of Theoretical Science, University of Oregon\\
5203 University of Oregon,
OR 97403-5203 Eugene, USA\\
E-mail: soper@physics.uoregon.edu
}
\maketitle

\abstracts{
In this paper we outline a new parton shower algorithm based on the Catani-Seymour dipole factorization. Our motivation is to have an algorithm which can naturally cooperate with the NLO calculations.
}

\section{Introduction}

One often uses perturbation theory to produce predictions for the results
of particle physics experiments in which the strong interaction is
involved. In order to get useable predictions one has to calculate at least at 
next-to-leading order to avoid large uncertainties those come from the unphysical scale dependences.  

Unfortunately, standard NLO programs have significant flaws. One flaw is 
that the final states consist just of a few partons, while in nature final states consist of many hadrons. A worse flaw is that the weights are often 
very large positive numbers or very large negative numbers. 

There is another class of calculational tools, the shower Monte Carlo
event generators, such as H{\small ERWIG} [\refcite{Herwig}] and P{\small
YTHIA} [\refcite{Pythia}].  These have the significant
advantage that the objects in the final state consist of hadrons.
Furthermore, the weights are never large numbers. Finally, the programs 
have a lot of important structure of QCD built into them. For at least some cases like this, the shower Monte Carlo programs can provide a good approximation for the cross sections. The chief disadvantage of typical shower Monte Carlo event generators is that they are based on leading order
perturbation theory for the basic hard process. This means that the predictions of the Monte Carlo programs usually have large theoretical uncertainty. From this reason we cannot use the Monte Carlo prediction as theoretical result in the experimental analysis. This rises an obvious question. Is it possible to build a Monte Carlo program that is predictable?
 
The parton distribution function or the fragmentation function is a convolution of a perturbative and a non-perturbative part. The perturbative part is the evolution kernel which is solution of the DGLAP equation. It is based on the collinear factorization property of QCD. Fixing the accuracy of the calculation and the factorization scheme the evolution kernel is well defined. The conventional factorization scheme is the $\overline{\rm MS}$. The non-perturbative part cannot be calculated, fortunately it is universal and we can obtain it from fit to the data and use them for other processes. Of course if we change the factorization scheme we have to refit the non-perturbative part.

What is the parton shower and the hadronization? 
We can think about that is a multi-hadron fragmentation function and the parton shower is its evolution kernel. The hadronization model is the non-perturbative part that we fit to the data. The perturbative part is based on the factorization theorem of the QCD and it is well defined up to the finite pieces. In other words, a different shower algorithm represent different {\em shower scheme} but upto the leading (LL) and next-to-leading (NLL) logarithmic accuracy  they must be equivalent. If we have some free parameters other than the trivial scales in the parton shower algorithm then the different values of these parameters correspond to different shower schemes and the tunning of the hadronization model must be redone. 

How to define a {\em conventional} shower scheme? These are our criteria:
i) It must be Lorentz covariant and Lorentz invariant. 
ii) It must be correct at LL level. It must be correct at NLL level at least in leading color appriximation. 
iii) The soft gluon effect must be correct at least in leading color approximation. 
iv) In every step of the shower the phase space configuration must be the exact $m$-body phase space with the exact phase space weight. The infrared cutoff parameter should be the only cutoff parameter in the algorithm.
v) It should ``smoothly'' work together with the NLO and matrix element matching schemes.

In this paper we aim to outline some new ideas for a parton shower algorithm along our criteria. In the next we will discuss a parton shower algorithm based on the Catani-Seymour dipole factorization formulas [\refcite{Catani:1996vz}] and give a general prescription for the matching of parton shower and fix order computation at leading and next-to-leading order level. It was a obvious choice 
to use the dipole factorization because it fulfills our criteria and it makes the NLO level matching easier because lots of modern NLO fix order computations are based on the dipole subtraction method, for example N{\small LOJET++} [\refcite{Nagy}] and M{\small CFM} [\refcite{MCFM}].

Note, this work is still in progress and we haven't published a proper paper about it. Since this is a workshop proceeding there is no way to define every details and proof of every statements precisely.

\section{Notation}

\subsection{Configuration space}

 In order to describe showers, we adopt a vector notation for classical
statistical physics. In order not to confuse classical states with
quantum states, we use the notation $|A)$ to denote a classical state $A$
instead of using $|A\rangle$, which we reserve for a quantum state. The
classical inner product is real, with $(A|B) =(B|A)$. We denote a state
consisting of two incoming partons $\La,\Lb$ and $m$ outgoing partons with momenta and flavours
\begin{equation}
\{p,f\}_{\La,\Lb,m}\equiv
\{\eta_{\La}p_{A}, a, \eta_{\Lb}p_{B}, b, p_{1}, f_{1}, ...,p_{m}, f_{m}\}
\end{equation}
by $|\{p,f\}_{\La,\Lb,m})$. We can use these as basis states, with normalization
\begin{equation}
\label{eq:normalization}
\begin{split}
(\{p',f'\}_{\La,\Lb, m'}|\{p,f\}_{\La,\Lb, m})={}&
 	\delta_{mm'}\, \delta_{a,a'}\,\delta(\eta_{\La}-\eta_{\La}')\, 
	\delta_{b,b'}\,\delta(\eta_{\Lb}-\eta_{\Lb}')\\
&\times  \prod_{i=1}^m \delta_{f,f_i'}\,\delta^{(4)}(p_i- p'_i)
\end{split}
\end{equation}
and completeness relation
\begin{equation}
\label{eq:completeness}
  1 = \sum_m \int \big[d\{p,f\}_{\La,\Lb, m}\big]\,
  |\{p,f\}_{\La,\Lb, m})(\{p,f\}_{\La,\Lb, m}|\;\;,
\end{equation}
where we introduced the an abbrevation for the integrals and sums, that is
\begin{equation}
\int \big[d\{p,f\}_{\La,\Lb, m}\big] \equiv \prod_{i=1}^m
\left\{\sum_{f_i}\int d^4p_i\right\}
\sum_{a}\int_{0}^{1}\!d\eta_{\La}
\sum_{b}\int_{0}^{1}\!d\eta_{\Lb}\;\;.
\end{equation}

We can use the basis states to express the differential probability $A(\{p,f\}_{\La,\Lb, m})$ that a general state $|A)$ consists of partons $\{p,f\}_{\La,\Lb, m}$:
\begin{equation}
  A(\{p,f\}_{\La,\Lb, m}) = (\{p,f\}_{\La,\Lb, m}|A)
\end{equation}
Then
\begin{equation}
\label{eq:physstate}
  |A) = \sum_m\int \big[d\{p,f\}_{\La,\Lb, m}\big]\,
  |\{p,f\}_{\La,\Lb, m})(\{p,f\}_{\La,\Lb, m}|A)\;\;.
\end{equation}
Since $(\{p,f\}_m|A)$ is a probability, a physical state $|A)$ must
obey the normalization condition $(1|A) = 1$, where $|1)$ is the vector 
with $(\{p,f\}_{\La,\Lb, m}|1) = 1$. More explicitly the unit vector is
\begin{equation}
\label{eq:unitstate}
(1| = \sum_m\int \big[d\{p,f\}_{\La,\Lb, m}\big]\,(\{p,f\}_{\La,\Lb, m}|\;\;.
\end{equation}

Given a measurement function $F$ for which the value measured in a parton
state $|\{p,f\}_{\La,\Lb, m})$ is $F(\{p,f\}_{\La,\Lb, m})$, it is convenient to define a vector $|F)$  by
\begin{equation}
\label{eq:Fmeasurevec}
  (F|\{p,f\}_{\La,\Lb, m}) = F(\{p,f\}_{\La,\Lb, m}).
\end{equation}
This is a convenient notation because then the expectation value of $F$
in a state $A$ is $\la F\ra_A = (F|A)$. 

\subsection{Description of color flow}

We will need a description of the color state that is adapted to a
description of shower evolution. If we use an index notation, to each
parton with label $l$ there is associated a color index $A_l$, which
takes values $1,\dots,3$ for a quark or antiquark and takes values
$1,\dots,8$ for a gluon. Thus the matrix element has color and
spin components and we can write
\begin{equation}
{\cal M}^{A_1,\dots,A_m}_{s_1,\dots,s_m}
  = \sum_{\{c\}_{m}} V(\{c\}_{m})^{A_1,\dots,A_m}
{\cal M}(\{p,f,c\}_{m})_{s_1,\dots,s_m} \;,
\end{equation}
where the $V(\{c\}_{m})$ form a basis for the color space with labels 
$\{c\}_{m}$ to be explained presently and the ${\cal M}(\{p,f,c\}_{m})$ are expansion coefficients (which are still vectors in spin space). In a vector notation, this is
\begin{equation}
|{\cal M}\rangle_{\rm c,s} = \sum_{\{c\}_{m}}
|V(\{c\}_{m})\rangle_{\rm c}
\otimes |{\cal M}(\{p,f,c\}_{m})\rangle_{\rm s}
\;.
\end{equation}
Here $|{\cal M}\rangle_{\rm c,s}$ lies in the combined color-spin space
while $|V(\{c\}_{m})\rangle_{\rm c}$ is a vector in color space and $|{\cal
M}(\{p,f,c\}_{m})\rangle_{\rm s}$ is a vector in spin space.

The color basis vectors $|V(\{c\}_{m})\rangle_{\rm c}$ are labelled by a {\it
color string configurations} $\{c\}_{m}$. We define a string configuration 
to be a set $\{S_1,\dots,S_n\}$ of one or more {\it strings} $S$. We
define strings to be of two types, {\it open strings} and {\it closed
strings}. We define an open string to be an ordered set of parton indices
$S = \{l_1,l_2,\dots,l_{n-1},l_n\}$, where $l_1$ is the label of an antiquark, 
$l_n$ is the label of a quark, and $l_2,\dots,l_{n-1}$ are labels of gluons. A 
closed string is an ordered set of parton indices 
$S = \{l_1,l_2,\dots,l_{n-1},l_n\}$, where all of the indices label gluons and
where we treat sets that differ by a cyclic permutation of the indices as being
the same.

We describe the color string configuration by assigning a color connection label $c_{i}$ to every external parton with label $i\in\{\La,\Lb,1,...,m\}$. For a gluon, $c_{i} = (\tilde\pi(i), \pi(i))$, for a quark $c_{i} = (0, \pi(i))$ and for an antiquark $c_{i} = (\tilde\pi(i), 0)$. Here $\pi(i)$ is the label of the next parton on the string while $\tilde\pi(i)$ is the label of the previous parton. This labeling respects that every open stings start with a quark and end with an antiquark. The allowed string configurations are those in which each parton index is an element of one of the strings.

Now we can define the basis states. We take $V(\{c\}_{m})$ to be a product
\begin{equation}
V(\{c\}_{m})^{A_1,\dots,A_m} =
V(S_1)^{\{A\}_{[1]}}\,
V(S_2)^{\{A\}_{[2]}}
\dots 
V(S_n)^{\{A\}_{[n]}}
\;.
\end{equation}
Here we have denoted the set of color indices represented in string $k$
by $\{A\}_{[k]} = \{A_{l_1},\dots,A_{l_n}\}$ if string $k$ is
$S_k = \{l_1,\dots, l_n\}$.

We can now define the component factors $V(S)$. We first consider an open
string. For notational convenience, we suppose that the partons along the
string are numbered sequentially $S = \{1,2,\dots,n\}$. With these labels 
for the partons, we define
\begin{equation}
V(S)^{\{A\}}
= \left[
t^{A_2}t^{A_3}\cdots t^{A_{n-1}}
\right]_{A_1A_n}
\;\;,
\end{equation}
where the $t^A$ are the SU(3) generator matrices for the fundamental
representation and we take the $A_1,A_n$ matrix element of the matrix
product of the generator matrices. For a closed string with the same
parton labels (now all gluons) we define
\begin{equation}
V(S)^{\{A\}}
= {\rm Tr}\left[
t^{A_1}t^{A_2}\cdots t^{A_{n}}
\right]
\;\;.
\end{equation}

A calculation of a cross section involves square of the matrix element. Since the basis in the color space is orthogonal up to the leading color terms for the matrix element square we have
\begin{equation}
\left|{\cal M}(\{p,f\}_{\La,\Lb,m})\right|^{2} = \!\!
\sum_{\{c\}_{\La,\Lb,m}}\!\!\!
\left|V(\{c\}_{\La,\Lb,m})\right|^{2}
\left|M(\{p,f,c\}_{\La,\Lb,m})\right|^{2}
+\cdots\;\;,
\end{equation}
where the dots stands for the subleading color contributions. They are suppressed by a factor of $1/N_{c}^{2}$. 

Each term in this sum represents a particular color configuration. Since the shower depends on the color configuration, from each term we generates a different shower. On the other hand, the leading color approximation is useful for nearly collinear parton configurations but is gives a poor description of the hard part of the event. For better accuracy we should begin the showering using the exact matrix element squared reweighted with the probability that the hard partons are in a certain color configuration. For a given color configuration $\{c\}_{\La,\Lb,m}$ this probability is 
\begin{equation}
p(\{p,f,c\}_{\La,\Lb,m}) = 
\frac{\left|V(\{c\}_{\La,\Lb,m})\right|^{2}\left|M(\{p,f,c\}_{\La,\Lb,m})\right|^{2}}
{\displaystyle
 \sum_{\{c'\}_{\La,\Lb,m}}\left|V(\{c\}_{\La,\Lb,m})\right|^{2}
 \left|M(\{p,f,c\}_{\La,\Lb,m})\right|^{2}}\;\;.
\end{equation}
Now the color configuration assigned matrix element square is 
\begin{equation}
\label{eq:colorassignME}
\left|{\cal M}(\{p,f,c\}_{\La,\Lb,m})\right|^{2} = 
p(\{p,f,c\}_{\La,\Lb,m})\,
\left|{\cal M}(\{p,f\}_{\La,\Lb,m})\right|^{2}\;\;,
\end{equation}
where $\left|{\cal M}(\{p,f\}_{\La,\Lb,m})\right|^{2}$ is the exact matrix element square.

It is useful to extend the configuration space by adding color information. Introducing the basis vector of the color space $|\{c\}_{\La,\Lb,m})$ the new configration space is defined as a tensor product that is
\begin{equation}
\sket{\{p,f,c\}_{\La,\Lb,m}}\equiv
\sket{\{p,f\}_{\La,\Lb,m}}\otimes\sket{\{c\}_{\La,\Lb,m}}
\end{equation}
and the normalization condition reads
\begin{equation}
\sbra{\{c\}_{\La,\Lb,m}} \{c'\}_{\La,\Lb,m'}\big) = 
\delta_{mm'}
\delta_{c_{\La}c'_{\La}}\delta_{c_{\Lb}c'_{\Lb}}
\prod_{l=1}^{m}\delta_{c_{l}c'_{l}}
\;\;.
\end{equation}
The integral over the state involves a sum over all possible color configurations
\begin{equation}
\int\big[d\{p,f,c\}_{\La,\Lb,m}\big]\equiv
\int\big[d\{p,f\}_{\La,\Lb,m}\big]
\sum_{\{c\}_{\La,\Lb,m}}\;\;.
\end{equation}
The $m$-parton matrix element square as a vector in the configuration space is
\begin{equation}
\label{eq:MEvector}
\sket{{\cal M}_{m}} = \int\big[d\{p,f,c\}_{\La,\Lb,m}\big]\,
\sket{\{p,f,c\}_{\La,\Lb,m}}
\, \left|{\cal M}(\{p,f,c\}_{\La,\Lb,m})\right|^{2}\;\;,
\end{equation}
where the color assigned amplitude $|{\cal M}(\{p,f,c\}_{\La,\Lb,m})|^{2}$ is defined in Eq.~\eqref{eq:colorassignME}.

\subsection{Phase space integrals and Born cross sections}

To define the phase space integral and include the parton distribution function it is useful to define an operator in the configuration space. Let us define 
\begin{equation}
\label{eq:opGamma}
\begin{split}
  \Gamma = \sum_{m}&\int\big[d\{p,f,c\}_{\La,\Lb,m}\big]\,
\sket{\{p,f, c\}_{\La,\Lb,m}}\sbra{\{p,f,c\}_{\La,\Lb,m}}
\\
&\!\times
f_{a/A}(\eta_{\La},\mu^{2}_{F})\,
f_{b/B}(\eta_{\Lb},\mu^{2}_{F})\,
\frac{1}{2\eta_{\La}\eta_{\Lb}p_{A}\!\cdot\!p_{B}}\,
\frac{1}{m!}
\\
&\!\times
\prod_{i=1}^{m}\left\{\frac{1}{(2\pi)^{3}}\delta_{+}(p_{i}^{2})\right\}\,
(2\pi)^{4}
\delta\bigg(\eta_{\La}p_{A}+\eta_{\Lb}p_{B}-K-\sum_{i=1}^{m}p_{i}\bigg)\;\;.
\end{split}
\end{equation}
With this definition we can easy define the phase space integral with the proper statistical ($1/m!$) and flux ($1/2\eta_{\La}\eta_{\Lb}p_{A}\!\cdot\!p_{B}$) factors. Here $K$ is the sum of the momenta of the non-QCD particles.  

Now we can define the $m$-parton Born level cross section as a vector in the configuration space
\begin{equation}
\label{eq:borncrossvec}
\big|\sigma_{m}\big) = \Gamma\sket{{\cal M}_{m}}
\end{equation}
where the vector $|{\cal M}_{m})$ is given in Eq.~\eqref{eq:MEvector}.
 
With this notation we can easily define of the Born level cross section of any $N$-jet quantity. Assuming that $F$ is a $N$-jet sensitive quantity then its cross section is 
\begin{equation}
\sigma[F] = \big(F\big|\sigma_{N}\big) 
=\sbra{F}\Gamma\sket{{\cal M}_{m}}\;\;,
\end{equation}
where $(F|$ is also a vector in the configuration space
\begin{equation}
\big(F\big| = \sum_{m=N}^{\infty}
\int\big[d\{p,f,c\}_{\La,\Lb,m}\big]\,
F\big(\{p\}_{\La,\Lb,m}\big)\,
\big(\{p,f,c\}_{\La,\Lb,m}\big|\;\;.
\end{equation}

\section{Parton shower evolution}

In this section we set up a quite general framework for describing a
parton shower. We use an evolution variable $t$ that starts at $0$ and runs
to $\infty$ and  represents someting like $\log(Q^2/k_{\perp}^2)$, where $Q$ is 
some hard scale. In principle, the shower evolves to $t = \infty$
but it actually stops at an infrared cutoff $t_{\rm f}$. The evolving
shower is represented by a state $|A(t))$ that begins with an initial
state $|A(t_0))$. The evolution is given by a linear operator $U(t',t)$,
with
\begin{equation}
|A(t)) = U(t,t_0)|A(t_0))\;\;.
\end{equation}
These operators have the group composition property
\begin{equation}
U(t_3,t_2)\, U(t_2,t_1)  = U(t_3,t_1)\;\;.
\end{equation}
The evolution operators preserve probabilities:
\begin{equation}
(1|U(t',t)|A) = (1|A)\;\;.
\end{equation}
for any state $|A)$. Here $(1|A) = 1$ if the state is normalized to unit
probability. 

The class of evolution operators that will use is defined by two
operators. The first is an infinitesimal generator of evolution or
hamiltonian, ${\cal H}(t)$, which can be specified by giving its matrix
elements
\begin{equation}
\sbra{\{p',f',c'\}_{\La,\Lb, m'}}{\cal H}(t)\sket{\{p,f,c\}_{\La,\Lb, m}}\;\;.
\end{equation}
The second operator is a no-splitting operator $N(t',t)$ that leaves the
basis states $|\{p,f,c\}_{\La,\Lb, m})$ unchanged except for multiplying each
of them by an eigenvalue $\Delta(\{p,f,c\}_{\La,\Lb, m};t',t)$:
\begin{equation}
N(t',t)\sket{\{p,f,c\}_{\La,\Lb, m}} =
\Delta(\{p,f,c\}_{\La,\Lb, m};t',t)\sket{\{p,f,c\}_{\La,\Lb, m}} \;\;.
\end{equation}
The evolution operator is expressed in terms of the hamiltonian and the
no-splitting operators by
\begin{equation}
\label{eq:evolution}
U(t_3,t_1) = N(t_3,t_1)
+ \int_{t_1}^{t_3}\! dt_2\ U(t_3,t_2)\,{\cal H}(t_2)\,N(t_2,t_1)
\;\;.
\end{equation}
This equation is interpreted as saying that either the system evolves
without splitting from $t_1$ to $t_3$, or else it evolves without
splitting until an intermediate time $t_2$, splits at $t_2$, and then
evolves (possibly with further splittings) from $t_2$ to
$t_3$.

There is a relation between the eigenvalues $\Delta$ of $N$ and the
matrix elements of ${\cal H}$. To derive this we apply the evolution operator 
for a basis state and using the normalization condition one can get  
\begin{equation}
\label{eq:sudakovgen}
\Delta(\{p,f,c\}_{\La,\Lb, m}; t_{2},t_{1}) = 
\exp\left(-\int_{t_1}^{t_{2}}\!dt\,
\sbra{1}{\cal H}(t)\sket{\{p, f,c\}_{\La,\Lb, m}}
\right)\;\;.
\end{equation}
Note that the probabilistic meaning of Eq.~\eqref{eq:sudakovgen} 
requires that the exponent be negative definite and strictly monotonic in the 
variable $t_{2}$.

One of the main purpose of the parton shower programs is to resum the leading and next-to-leading logarithms which come from the soft and collinear regions. In this phase space regions the matrix elements have factorization property that 
any $m+1$ parton matrix element can be written as a product of a singular factor 
and the corresponding $m$ parton matrix element.  It is known that the leading 
and next-to-leading logarithm are produced by the $1\to2$ type splittings. 
Thus it is enough to consider only the $1\to2$ splittings in the splitting operator ${\cal H}(t)$ and we have 
\begin{equation}
\begin{split}
  \sbra{1}{\cal H}(t)\sket{\{p, f,c\}_{\La,\Lb, m}} =
\int &\big[d\{\hat p,\hat f,\hat c\}_{\La,\Lb, m+1}\big]\,
\\
&\times\sbra{\{\hat p,\hat f,\hat c\}_{\La,\Lb, m+1}}{\cal H}(t)
\sket{\{p, f,c\}_{\La,\Lb, m}}\;\;.
\end{split}
\end{equation}

This is a very formal definition of the shower algorithm. Let us try to define the splitting operators. 

\section{Splitting operators}

The formalism outlined above can be used to describe showering with a variety of splitting operators. The definition of the splitting operators tht we suggest  is based on the Catani-Seymour dipole factorization formulas and is given by
\begin{equation}
\label{eq:splitoper}
\begin{split}
\big(\{&\hat p,\hat f,\hat c\}_{\La,\Lb, m+1}\big|{\cal H}(t)
\sket{\{p, f, c\}_{\La,\Lb, m}} 
\\
& = 
\sum_{\substack{l,k=\La,\Lb, 1,...,m\\ k\neq l}}
\int_{0}^{1}\!\frac{dy}{y}
\int_{0}^{1}\!dz \int_{0}^{2\pi} \frac{d\phi}{2\pi}\,
\delta\big(t + \log(T_{l,k}(p_{l}, p_{k}, z, y)/Q^{2})\big)
\\
&\quad\times
\frac{\as\big(Q^{2}e^{-t}\big)}{2\pi}\,
S_{l,k}(z,y, \hat f_{l,1}, \hat f_{l,2})\,
\frac{\hat\eta_{\La}}{\eta_{\La}}
\frac{f_{\hat a/A}(\hat\eta_{\La}, \mu_{F}^{2})}
{f_{a/A}(\eta_{\La}, \mu_{F}^{2})}\,
\frac{\hat\eta_{\Lb}}{\eta_{\Lb}}
\frac{f_{\hat b/B}(\hat\eta_{\Lb}, \mu_{F}^{2})}
{f_{b/B}(\eta_{\Lb}, \mu_{F}^{2})}\,
\\
&\quad\times A_{f_{l}}
\chi_{l,k}(\{c\}_{\La,\Lb, m})\, 
%\sbra{\{\hat c\}_{\La,\Lb, m+1}}
%{\cal T}_{l,k}(\hat f_{l,1}, \hat f_{l,2})
%\sket{\{c\}_{\La,\Lb, m}}
%\\
%&\qquad\times
\sbra{\{\hat p,\hat f, \hat c\}_{\La,\Lb, m+1}}
{\cal R}_{l,k}(z,y,\kappa_{\perp})
\sket{\{p, f, c\}_{\La,\Lb, m}}
\;\;.
\end{split}
\end{equation}
On the right hand side we have a sum over all the possible emitters 
($l\in\{\La,\Lb, 1,...,m\}$). In the general case when we have initial state hadrons the incoming partons are included among the allowed emitters. There is a sum also over the possible spectator partons ($k \in \{\La,\Lb, 1,...,m\}$ with $k\neq l$). The spitting is parametrized by the splitting variables $y$, $z$ and $\phi$. The variable $y$ is a virtuality like variable, $z$ is a momentum fraction variable and $\phi$ is an azimuthal angle that parametrizes the space-like unit vector $\kappa_{\perp}$ ($\kappa^{2}_{\perp} = -1$) which is perpendicular to the emitter and spectator momenta.

The dimensionless evolution variable $t$ is given by the function $T_{l,k}$ and depends on the $\{p, f\}_{\La,\Lb, m}$ configuration and on the splitting variables $y$, and $z$. This function could be any fully infrared sensitive variable. We prefer to use the transverse momentum of the emitted partons as defined in the centre-of-mass frame of the emitter and spectator. The transverse momentum is perpendicular to the momenta of both emitter and spectator. With this definition, it is boost invariant when we go back to the original frame.
The strong coupling also calculated at the evolution scale.

The the splitting kernel is given by the the function 
$S_{l,k}(z,y, \hat f_{l,1}, \hat f_{l,2})$. It is derived from the soft and 
collinear behavior of the tree level matrix elements and related to the Altarelli-Parisi splitting functions. In the case when we have initial state splitting then according to the backward evolution the flavor  $\hat f_{\La,1}\equiv\hat a$ is the flavor of  the new initial state parton. 
The spectator can be either a final state or an initial state parton and of course the functional form of the splitting kernel can depend on the type of the spectator parton.
Since the functional form of the splitting function depends on the emitter and spectator, one can distinguish four cases. They are
\begin{equation}
S_{l,k}(z,y, \hat f_{l,1}, \hat f_{l,2}) = 
\begin{cases}
S_{\rm fin}(z,y, \hat f_{l,1}, \hat f_{l,2})
\quad&\text{if}\;\;l,k\in\{1,...,m\}
\\
\tilde S_{\rm fin}(z,y, \hat f_{l,1}, \hat f_{l,2})
\quad&\text{if}\;\;l\in\{1,...,m\}\;\;\&\;\;k\in\{\La,\Lb\}
\\
S_{\rm ini}(z,y, \hat f_{l,1}, \hat f_{l,2})
\quad&\text{if}\;\;l\in\{\La,\Lb\}\;\;\&\;\;k\in\{1,...,m\}
\\
\tilde S_{\rm ini}(z,y, \hat f_{l,1}, \hat f_{l,2})
\quad&\text{if}\;\;l,k\in\{\La,\Lb\}
\end{cases}
\end{equation}

When the emitter or the spectator parton is one of the initial state partons, the momentum fraction of the incoming parton changes after the splitting. This effect is reflected by the the ratios of the parton distribution functions in Eq.~(\ref{eq:splitoper}). The parton distribution functions depend on the arbitrary factorization scale $\mu^{2}_{F}$. In a shower program $\mu^{2}_{F}$ is usually set to the evolution variable. However, the exact choice affects only subleading logarithms.

The operator ${\cal R}_{l,k}$ in Eq.~\eqref{eq:splitoper} describes the 
$\{p, f,c\}_{\La,\Lb, m}\to\{\hat p,\hat f, \hat c\}_{\La,\Lb, m+1}$ transformation and provides the correspondent Jacobians.

In the dipole formalism, a splitting always involves two partons. One is the emitter (labeled by $l$ in Eq.~\eqref{eq:splitoper}) that splits and produces two daugther partons. The other is the spectator (labeled by $k$ in Eq.~\eqref{eq:splitoper}) which absorbs the recoiled momentum and plays an important role in the color dynamics. With this construction, one ensures that the momenta are on-shell and that momentum conservation is maintained exactly. Furthermore, the splitting probability includes the exact phase space weight. 

How should one choose the spectator parton in a shower picture? In keeping with the role of the spectator in the color dynamics for a soft emission in the Catani-Seymour scheme, the spectator parton should be color connected to the emitter parton. In order to have a good approximation in the soft limit we should include the color connection between the emitter and spectator 
($-\bm{T}_{l}\cdot\bm{T}_{k}/\bm{T}_{l}^{2}$). Expanding this operator in 
$1/N_{c}$ then we have
\begin{equation}
\frac{\bm{T}_{l}\cdot\bm{T}_{k}}{-\bm{T}_{l}^{2}}
= A_{f_{l}}\,\chi_{l,k}\!\left(\{c\}_{\La,\Lb,m}\right)
+ {\cal O}\!\left(\frac{1}{N_{c}^{2}}\right)\;\;,
\end{equation}
where factor $A_{f_{l} = g} = 1/2$ for a gluon emitter and for quark or antiquark emitter it is  $A_{f_{l} = q,\bar q} = 1$. The function $\chi_{l,k}$
is
\begin{equation}
\chi_{l,k}\!\left(\{c\}_{\La,\Lb,m}\right) = 
\begin{cases}
1 \quad&\text{if}\;\; c_{l} = (k,\pi(l))\;\;\text{or}\;\;c_{l} = (\pi'(l), k)\\
0       &\text{otherwise}
\end{cases}\;\;\;\;.
\end{equation}
Remember, our color basis are based on open and closed color strings. The spectator parton is one of the neighbor parton of the emitter on the color string. When the gluon is the emitter there are two possible spectator because the gluon always has two neighbors on the color sting while for quarks and antiquarks there is always one possible spectator since they are always at the end of the color string. 

\subsection{Final state splitting with final state spectator}

Let see an example for the splitting functions and for the transformations. 
In the case when both the emitter and the spectator are final state partons 
($l,k\in\{1,...,m\}$) the splitting function is given by
\begin{equation}
\label{flavorDipoleFF:S}
\begin{split}
S_{\rm fin}(z,y, f_{1}, f_{2}) ={}&
\sum_{r=u,\bar u,d,\bar d,\dots}\!\!\biggl[
      \delta_{rf_1}\delta_{\bar{r}f_2}
      \, S_{q\bar{q}}(z,y)
      +\delta_{rf_1}\delta_{gf_2}\,S_{qg}(z,y)\\
      &\qquad\qquad\quad
      + \delta_{gf_1}\delta_{rf_2}\, S_{qg}(1-z,y)\biggr] 
\\  &
+ \delta_{gf_1}\delta_{gf_2}\, S_{gg}(z,y)
\;\;.
\end{split}
\end{equation}

Then for the splitting of a quark or antiquark into the same flavor quark
or antiquark plus a gluon the splitting function is  
\begin{equation}
  \label{DipoleFF:Sqg}
  S_{qg}(z, y) = 
    C_{\rm F}\,\left[\frac2{1-z(1-y)} -(1+z)\right]\;\;.
\end{equation}
For the splitting of a gluon into a quark and an antiquark, we define
\begin{equation}
  \label{DipoleFF:Vqqb}
  S_{q\bar{q}}(p,z,y) 
  = T_{\rm R}\,\left[1- 2z(1-z)\right] \;\;.
\end{equation} 
Finally for the splitting of a gluon into two gluons the splitting
function is
\begin{equation}
\label{DipoleFF:Vgg}
S_{gg}(z,y) = 2C_{\rm A}\,
    \bigg[ \frac1{1-z(1-y)}+\frac1{1-(1-z)(1-y)}-2 + z(1-z)\bigg]
\;\;.
\end{equation}

The momentum and flavor transfomation is given by the matrix element of the  
${\cal R}_{l,k_{l}}$ operator and it is 
\begin{equation}
\begin{split}
\big(\{\hat p,\hat f&\}_{\La,\Lb, m+1}\big|
{\cal R}_{l,k}(z,y,\kappa_{\perp})
\sket{\{p, f\}_{\La,\Lb, m}}
\\
={}& 
\frac12(1-y)\,\delta_{\hat f_{l,1}+\hat f_{l,2}}^{f_l}\,
\delta_{\hat a}^{a}\,\delta(\hat\eta_{\La}-\eta_{\La})\,
\delta_{\hat b}^{b}\,\delta(\hat\eta_{\Lb}-\eta_{\Lb})\,
\prod_{\substack{i=1\\i \neq l}}^{m}\delta_{\hat f_i}^{f_i}
\\
&\times
\delta^{(4)}(\hat p_{k} - (1-y)p_{k})\,
\prod_{\substack{i=1\\i \neq l, k}}^{m}
\delta^{(4)}(\hat p_{i} - p_{i}) 
\\
&\times\delta^{(4)}\big(\hat p_{l,1} - zp_{l} - y(1-z)p_{k} 
- [2p_{l}\!\cdot\!p_{k}yz(1-z)]^{1/2}\kappa_{\perp}\big)
\\
&\times\delta^{(4)}\big(\hat p_{l,2} - (1-z)p_{l} - yzp_{k} 
+ [2p_{l}\!\cdot\!p_{k}yz(1-z)]^{1/2}\kappa_{\perp}\big)\;\;.
\end{split}
\end{equation}

The evolution variable is the magnitude of the transverse momentum, that is 
\begin{equation}
T_{l,k}(p_{l}, p_{k}, z, y) = 2p_{l}\!\cdot\!p_{k}\,y\,z(1-z)\;\;.
\end{equation}

\subsection{Including heavy quark contributions and SUSY-QCD}

The mass of the top quark and the strongly interacting SUSY particles is very big thus the probability of their production is low. In this work we don't want to deal with this cases because the phase space requires a more complicated treatment. In the future we want to extend our algorithm for the heavy and SUSY particles.

\section{\label{sec:showerxsec}Shower cross sections}

In order to calculate the shower cross sections in shower approximation we have to define the physical state at the starting (hard) scale $t_{0}$. The hardest part of the event is always the simplest process which is kinematically possible. For example in $e^{+}e^{-}$ annihilation the simplest process is $e^{+}e^{-}\to q\bar q$, or for Higgs production in proton-proton collision it is $pp\to H$.

For instance, if we want to study jet production in proton antiproton collision then the simplest process is the $2\to2$ QCD processes and $m = 2$. The $t_{0}$ is given by the hard part which could be the the invariant mass of the outgoing hard partons. The physical state at the starting scale is the simplest state that is kinematically possible
\begin{equation}
\big|\sigma(t_{0})\big) = \big|\sigma_{2}\big)\;\;,
\end{equation}
where $|\sigma_{2})$ is given in Eq.~\eqref{eq:borncrossvec}.
The shower cross section is just the evolution of the initial state
\begin{equation}
\label{eq:showerxsec}
\big|\sigma(t_{\rm f})\big) = 
U(t_{\rm f}, t_{0})\big|\sigma(t_{0})\big) \;\;, 
\end{equation}
where the $t_{\rm f}$ scale is the infrared cutoff scale. And the cross section of any jet quantity is given by $F$ is 
\begin{equation}
\sigma[F] = \big(F\big|D(t_{\rm f})\big|\sigma(t_{\rm f})\big) = 
\big(F\big|D(t_{\rm f}) U(t_{\rm f}, t_{0})\big|\sigma(t_{0})\big) \;\;,
\end{equation}
where the operator $D(t_{f})$ represents the hardornization. 

The hadronization is a long distance effect what we cannot calculate from QCD  but we have some QCD motivated model for it. In principle the model is universal (as the hadronization). Once the model has been fit to a set of the data ({\em e.g.} LEP data) we don't have to retune the model for another type of the processes. Although the hadronization model is universal, it is strongly coupled with the shower evolution. That means if we change the shower evolution we have to retune the hadronization model.

\section{Adjoint splitting operator}

The whole parton shower idea is based on the factorization properties of the matrix elements in the soft and collinear regions. In this region an $m+1$ parton  matrix element can be written as product an $m$-parton matrix element and an universial singular factor. Using this approximtion, one can start from a very simple configuration and generate large multiplicity multiparton states.

In our formalism it is essential that the phase space is the exact $m$-body phase space with the exact weight in every step of the shower. Furthermore the shower is Lorentz invariant. We saw that applying the operator ${\cal H}(t)$ to an $m$-parton state, one obtains an $m+1$ parton state. The probability for the  emission is the $m+1$ parton phase space weight times a universal singular factor. 

Now we can ask if it is possible to do this other way around. Suppose that we have generated an $m+1$ parton phase space point. Can we define an operator that acts on this state and it finds all the possible $m$-parton states with corresponding weights such that these would be the weights of an emission if the operator ${\cal H}$ was applied on these $m$-parton states. We claim it is possible to define this operator since the factorization works exactly this way. 

Let us define the adjoint splitting operator ${\cal H}^{\dagger}(t)$ that obeys to the following condition:
\begin{equation}
\label{eq:adjHdef}
\sbra{F}{\cal H}(t)\Gamma\sket{A} = 
\sbra{A}{\cal H}^{\dagger}(t)\Gamma\sket{F}\;\;,
\end{equation}
where the states $|A)$ and $|F)$ are arbitrary and the operator $\Gamma$ makes the integrals phase space integrals as it is defined in Eq.~\eqref{eq:opGamma}.
We can see immediately that the operator ${\cal H}^{\dagger}(t)$ always decreases the number of partons because ${\cal H}(t)$ always increases it. Since in ${\cal H}(t)$ we consider only $1\to2$ splittings in ${\cal H}^{\dagger}(t)$ we have only $2\to1$ merging. Thus there is only one non-zero matrix element   
\begin{equation}
\label{eq:adjont1H}
\begin{split}
  \sbra{1}{\cal H}^{\dagger}&(t)\sket{\{p, f,c\}_{\La,\Lb, m}} 
  \\
  &=
\int\big[d\{\tilde p,\tilde f, \tilde c\}_{\La,\Lb, m-1}\big]
\sbra{\{\tilde p,\tilde f, \tilde c\}_{\La,\Lb, m-1}}
{\cal H}^{\dagger}(t)\sket{\{p, f,c\}_{\La,\Lb, m}}\;\;.
\end{split}
\end{equation}
From Eq.~\eqref{eq:adjHdef} we can immediately see that for several emission we can write
\begin{equation}
\label{eq:adjHstring}
\sbra{F}{\cal H}(t_{m}){\cal H}(t_{m-1})\cdots{\cal H}(t_{3})\Gamma\sket{A} 
= 
\sbra{A}{\cal H}^{\dagger}(t_{3})\cdots{\cal H}^{\dagger}(t_{m-1})
{\cal H}^{\dagger}(t_{m})\Gamma\sket{F}\;\;.
\end{equation}

Using Eq.~\eqref{eq:splitoper} and from the Catani-Seymour dipole subtraction terms [\refcite{Catani:1996vz}] for the non-zero matrix elements of operator ${\cal H}^{\dagger}(t)$ we have 
\begin{equation}
\label{eq:adjsplitoper}
\begin{split}
\big(\{\tilde p,\tilde f,\tilde c&\}_{\La,\Lb, m}\big|{\cal H}^{\dagger}(t)
\sket{\{p, f,c\}_{\La,\Lb, m+1}} 
\\
&= 
\sum_{\substack{i,j\\ \rm pairs}}\sum_{k\neq i,j}
\frac{1}{2p_{i}\!\cdot\!p_{j}}\,
\frac{\eta_{\La}}{\tilde\eta_{\La}}
\frac{\eta_{\Lb}}{\tilde\eta_{\Lb}}\,
\langle\bm{V}_{ij,k}(p_{i}, p_{j}, p_{k}, f_{i}, f_{j})\rangle\,
\\
&\qquad\times 
A_{f_{ij}}\chi_{ij,k}(\{\tilde c\}_{\La,\Lb, m})\,
\delta\big(t + \log(T_{ij,k}(p_{i}, p_{j}, p_{k})/Q^{2})\big)\\
&\qquad\times
\sbra{\{\tilde p,\tilde f,\tilde c\}_{\La,\Lb, m}}
{\cal Q}_{ij,k}\sket{\{p, f,c\}_{\La,\Lb, m+1}}
\;\;.
\end{split}
\end{equation}

Here we have sum first over all the possible pairs $\{i,j\}$ of partons that could be the daughter partons from the splitting. The daughter partons must be color connected. Then we sum over possible choices $k$ of the spectator parton. 
The Dirac delta function maps out the dimensionless evolution variable $t$, which is the logarithm of the transverse momentum $T_{ij,k}(p_{i}, p_{j}, p_{k})$ in the splitting. We have the spin averaged splitting function 
$\langle\bm{V}_{ij,k}\rangle$ that depends on the momenta of partons $i,j,k\in\{\La,\Lb, 1,..m+1\}$ and the flavors of partons $i$ and $j$. It contains a 
factor of the strong coupling $\as(k_{\perp}^{2})$. The color connection between the emitter and spectator ($-\bm{T}_{ij}\cdot\bm{T}_{k}/\bm{T}^{2}_{ij}$) is considered only at leading color level, that is $A_{f_{ij}}\chi_{ij,k}$. 
The momentum, flavor and color transformation prescription is specified 
by the operator ${\cal Q}_{ij,k}$.

\section{Matching parton shower and Born level matrix elements} 

In Section \ref{sec:showerxsec} we defined the shower cross section as the shower evolution of the kinematically simplest hard process, for example the $2\to2$ processes in hadron-hadron collisions. It is obvious that this approximation is very poor if we are interested in the cross sections other than the 2-jet cross section. In order to have a better approximation one should consider higher multiplicity exact matrix elements.

Let us start with the shower cross section expand the first $n$ steps of the evolution operator 
\begin{equation}
\label{eq:showerexpand}
\begin{split}
\big|\sigma(t_{\rm f}&)\big) = 
N(t_{\rm f},t_{2})\sket{\sigma_{2}}\,
\\
&+ \sum_{m=3}^{n-1}
\int_{t_{2}}^{t_{\rm f}}\!\!dt_{3}\int_{t_{3}}^{t_{\rm f}}\!\!dt_{4}\,
\cdots\int_{t_{m-1}}^{t_{\rm f}}\!\!dt_{m}\,
N(t_{\rm f},t_{m}){\cal H}(t_{m})\,
\\
&\qquad\quad\times
N(t_{m},t_{m-1}){\cal H}(t_{m-1})\cdots
N(t_{4},t_{3}){\cal H}(t_{3})\,N(t_{\rm 3},t_{2})\sket{\sigma_{2}}
\\
&+ \int_{t_{2}}^{t_{\rm f}}\!\!dt_{3}\int_{t_{3}}^{t_{\rm f}}\!\!dt_{4}\,
\cdots\int_{t_{n-1}}^{t_{\rm f}}\!\!dt_{n}\,
U(t_{\rm f}, t_{n}){\cal H}(t_{n})\,
\\
&\qquad\quad\times
N(t_{n},t_{n-1}){\cal H}(t_{n-1})\cdots
N(t_{4},t_{3}){\cal H}(t_{3})\,N(t_{3},t_{2})\sket{\sigma_{2}}\;\;.
\end{split}
\end{equation}
In every term we have a string of operator products $H(t)N(t,t')$, which provide the probability of non-emission between ``times'' $t'$ and $t$ followed by a splitting at $t$.  This corresponds to the approximate matrix element squared times the proper Sudakov factors. We want to replace the approximate matrix element with the exact $m$-parton tree level matrix element while keeping the proper Sudakov weights.

We use the approximation of the matrix elements in the soft and collinear regions when emissions are ordered. One can show if the state 
$|\{p,f,c\}_{\La,\Lb,m})$ was generated according to the shower procedure 
then the following is a good approximation
\begin{equation}
\begin{split}
  \big({\cal M}_{m}&\sket{\{p,f,c\}_{\La,\Lb,m}}
\approx{}
 \big({\cal A}_{m}(t_{\rm f}, t_{2})\sket{\{p,f,c\}_{\La,\Lb,m}}
 \\
&\equiv
\int_{t_{2}}^{t_{\rm f}}\!\!dt_{3}\int_{t_{3}}^{t_{\rm f}}\!\!dt_{4}
\cdots\int_{t_{m-1}}^{t_{\rm f}}\!\!dt_{m}\,
\prod_{k=3}^{m}\frac{\as(\mu_{R}^{2})}{\as(Q^{2}e^{-t_{k}})}\,
\\
&\qquad
\times
\sbra{{\cal M}_{2}}{\cal H}^{\dagger}(t_{3}){\cal H}^{\dagger}(t_{4})
\cdots{\cal H}^{\dagger}(t_{m})\sket{\{p,f,c\}_{\La,\Lb,m}}\;\;.
\end{split}
\end{equation}
It is clear that knowing the state $|\{p,f,c\}_{\La,\Lb,m})$, the approximate matrix element $({\cal A}_{m}(t_{\rm f}, t_{2})|\{p,f,c\}_{\La,\Lb,m})$ is calculable according to the  definition of operator ${\cal H}^{\dagger}(t)$. However this could be a very big sum with lots of terms. One can immediately see that this approximation is valid only in those regions where the emissions are ordered and the phase space is dominated by soft and collinear radiation.
 
In the next step we reweight every term in Eq.~\eqref{eq:showerexpand} by a matrix element factor. Whenever $({\cal M}_{m}|$ is known and the state 
$|\{p,f,c\}_{\La,\Lb,m})$ is generated by the shower procedure, the following ratio is always well defined:
\begin{equation}
w_{M}(\{p,f,c\}_{\La,\Lb,m}, t_{\rm f}, t_{2})=
\frac{\displaystyle\big({\cal M}_{m}\sket{\{p,f,c\}_{\La,\Lb,m}}}
{\displaystyle \big({\cal A}_{m}(t_{\rm f}, t_{2})\sket{\{p,f,c\}_{\La,\Lb,m}}}
\;\;.
\end{equation}
When $|{\cal M}_{m})$ is unknown, we can simply set $w_{M}(\{p,f,c\}_{\La,\Lb,m}, t_{\rm f}, t_{2})=1$. Let us define the matrix element reweighting operator
\begin{equation}
\begin{split}
W_{M}(t_{\rm f}, t_{2}) = \sum_{m}
&\int\big[d\{p,f,c\}_{\La,\Lb,m}\big]\,
\sket{\{p,f,c\}_{\La,\Lb,m}}
\sbra{\{p,f,c\}_{\La,\Lb,m}}
\\
&\;\;\times
w_{M}(\{p,f,c\}_{\La,\Lb,m}, t_{\rm f}, t_{2})
\;\;.
\end{split}
\end{equation}
Acting on an $m$-parton state generated by the shower algorithm, this operator replaces the approximate squared matrix element by the exact one (if $m$ is small enough that the squared matrix element is known). Acting on a state with larger $m$ it acts as the unit operator. Using this reweighting operator, the improved shower cross section is 
\begin{equation}
\label{eq:showerexpandM}
\begin{split}
\big|\sigma_{M}(t_{\rm f})\big) = {}&
N(t_{\rm f},t_{2})\sket{\sigma_{2}}\,
\\
&+ \sum_{m=3}^{n-1}
\int_{t_{2}}^{t_{\rm f}}\!\!dt_{3}\int_{t_{3}}^{t_{\rm f}}\!\!dt_{4}\,
\cdots\int_{t_{m-1}}^{t_{\rm f}}\!\!dt_{m}\,
N(t_{\rm f},t_{m})W_{M}(t_{\rm f}, t_{2})
\\
&\qquad\qquad\times
{\cal H}(t_{m})N(t_{m},t_{m-1})\cdots
{\cal H}(t_{3})\,N(t_{\rm 3},t_{2})\sket{\sigma_{2}}
\\
&+ \int_{t_{2}}^{t_{\rm f}}\!\!dt_{3}\int_{t_{3}}^{t_{\rm f}}\!\!dt_{4}\,
\cdots\int_{t_{n-1}}^{t_{\rm f}}\!\!dt_{n}\,
U(t_{\rm f}, t_{n})W_{M}(t_{\rm f}, t_{2})
\\
&\qquad\qquad\times
{\cal H}(t_{n})N(t_{n},t_{n-1})\cdots
{\cal H}(t_{3})\,N(t_{3},t_{2})\sket{\sigma_{2}}\;\;.
\end{split}
\end{equation}
Here it is assumed that the matrix elements $|{\cal M}_{2}),\dots,|{\cal M}_{n})$ are known. The nice feature of this formulae that the event is generated according to a shower algorithm starting form the trivial $2\to2$ kinematics and whenever the exact $m$-parton matrix element is known the event is rewieghted by the ratio of the this tree level amplitude and the approximate shower amplitude. This form is suitable for numerical implementation. 

With the exact matrix element correction we don't change the leading and next-to-leading logarithms of the cross section. It is easy to see if we rewrite Eq.~\eqref{eq:showerexpandM} in an equivalent form by adding and subtracting the same terms. Thus we have
\begin{equation}
\label{eq:showerexpandMsub}
\begin{split}
\big|\sigma_{M}(&t_{\rm f})\big) = 
U(t_{\rm f},t_{2})\sket{\sigma_{2}}\,
\\
&+ \sum_{m=3}^{n}
\int_{t_{2}}^{t_{\rm f}}\!\!dt_{3}\int_{t_{3}}^{t_{\rm f}}\!\!dt_{4}\,
\cdots\int_{t_{m-1}}^{t_{\rm f}}\!\!dt_{m}\,
U(t_{\rm f},t_{m})
\big[W_{M}(t_{\rm f}, t_{2}),\, {\cal H}(t_{m})\big]
\\
&\qquad\qquad\times
N(t_{m},t_{m-1}){\cal H}(t_{m-1})\cdots
{\cal H}(t_{3})\,N(t_{\rm 3},t_{2})\sket{\sigma_{2}}
\end{split}
\end{equation}
and the commutator is defined in the usual way, that is 
\begin{equation}
\big[W_{M}(t_{\rm f}, t_{2}),\, {\cal H}(t_{m})\big]
=W_{M}(t_{\rm f}, t_{2}){\cal H}(t_{m})
-{\cal H}(t_{m})W_{M}(t_{\rm f}, t_{2})\;\;.
\end{equation}
In Eq.~\eqref{eq:showerexpandMsub} in the terms with higher multiplicity matrix element correction the leading (LL) and next-to-leading logarithms (NLL) are removed by the commutator leaving only subleading logarithms and finite pieces. All the LL and NLL logarithms are accumulated in the the first term which is actually the standard shower.

\subsection{Matching with Sudakov reweighting}

In this subsection we rewrite our matching formulae in an equivalent form. This form is less efficient from the point of the numerical implementation but it is very useful for further studies.
 
After some algebraic manipulation one can find that Eq.~\eqref{eq:showerexpandM} can be written in the following form:
\begin{equation}
\label{eq:showerexpandDelta}
\begin{split}
\big|\sigma_{M}(&t_{\rm f})\big) = 
\big|\sigma_{\Delta}(t_{\rm f})\big) \\
&\equiv 
N(t_{\rm f},t_{2})\sket{\sigma_{2}}\,
+ \sum_{m=3}^{n-1}
\int_{t_{2}}^{t_{\rm f}}\!\!dt_{m}\,
N(t_{\rm f},t_{m})W_{\Delta}(t_{\rm f}, t_{m}, t_{2})\sket{\sigma_{m}}
\\
&
+ \int_{t_{2}}^{t_{\rm f}}\!\!dt_{n}\,
U(t_{\rm f}, t_{n})W_{\Delta}(t_{\rm f}, t_{n}, t_{2})\sket{\sigma_{n}}\;\;,
\end{split}
\end{equation}
where the state $|\sigma_{m})$ is defined in Eq.~\eqref{eq:borncrossvec} and the Sudakov reweighting operator is
\begin{equation}
\label{eq:sudakovreweight}
\begin{split}
W_{\Delta}(t_{\rm f}, &t, t_{2}) =
\sum_{m}
\int\big[d\{p,f,c\}_{\La,\Lb,m}\big]\,
\sket{\{p,f,c\}_{\La,\Lb,m}}
\sbra{\{p,f,c\}_{\La,\Lb,m}}
\\
&\;\times
\int_{t_{2}}^{t}dt_{m-1}  
\int_{t_{2}}^{t_{m-1}}dt_{m-2}\cdots  
\int_{t_{2}}^{t_{4}}dt_{3}  
\\
&\qquad\times
\frac{\sbra{{\cal M}_{2}}N(t_{\rm 3},t_{2}){\cal H}^{\dagger}(t_{3})\cdots
N(t,t_{m-1}){\cal H}^{\dagger}(t)\sket{\{p,f,c\}_{\La,\Lb,m}}}
{\displaystyle \big({\cal A}_{m}(t_{\rm f}, t_{2})\sket{\{p,f,c\}_{\La,\Lb,m}}}
\;\;.
\end{split}
\end{equation}
This operator provides all the possible emission history with the proper Sudakov factor.

\subsection{Slicing method}
 
The slicing method was originally defined by Catani, Krauss, Kuhn and Webber 
[\refcite{Catani:2001cc}] for $e^{+}e^{-}$ annihilation and was later was 
generalized for processes with initial state partons by Krauss 
[\refcite{Krauss:2002up}] and Mrenna and Richardson [\refcite{Mrenna:2003if}]. 
Nowadays this method is very popular and one can find several implementations in the literature [\refcite{implckkw}]. It is usually called the CKKW-method. The basic idea is to divide the evolution region two parts, a hard part and a soft part. In the hard region one uses the exact matrix element supplemented by an appropriate Sudakov factor. The soft part is dominated by soft and collinear radiation so the shower cross section provides a good approximation.

Defining the matching scale $t_{\rm f} > t_{\rm ini} > t_{0}$ and  using the group decomposition property of the evolution operator then we have 
\begin{equation}
\big|\sigma(t_{\rm f})\big) = 
U(t_{\rm f}, t_{\rm ini})U(t_{\rm ini}, t_{2})\big|\sigma_{2}(t_{2})\big) \;\;. 
\end{equation}
Now, we can apply the matching formulae for 
$U(t_{\rm ini}, t_{2})|\sigma_{2}(t_{2}))$ according to Eq.~\eqref{eq:showerexpandDelta}. The CKKW-method uses a further approximation for the Sudakov reweighting operator. Instead of using \eqref{eq:showerexpandDelta}, the emission history and the Sudakov factors are defined according to the 
$k_{\perp}$ jet finding algorithm. With this method, a unique emission history is associated with each given $n$-parton configuration. For the CKKW cross section we have
\begin{equation}
\label{eq:showerCKKW}
\begin{split}
\big|\sigma_{\rm CKKW}&(t_{\rm f})\big) =
U(t_{\rm f}, t_{\rm ini})N(t_{\rm ini},t_{2})\sket{\sigma_{2}}\,
\\
&+ \sum_{m=3}^{n-1}
\int_{t_{2}}^{t_{\rm ini}}\!\!dt_{m}\,
U(t_{\rm f}, t_{\rm ini})N(t_{\rm ini},t_{m})
W_{\rm CKKW}(t_{\rm ini}, t_{m}, t_{2})
\sket{\sigma_{m}}
\\
&+ \int_{t_{2}}^{t_{\rm ini}}\!\!dt_{n}\,
U(t_{\rm ini}, t_{n})
W_{\rm CKKW}(t_{\rm ini}, t_{n}, t_{2})
\sket{\sigma_{n}}\;\;.
\end{split}
\end{equation}

The CKKW method has the advantage that the Sudakov reweighting is very simple. Thus the method is easy to implement. However, we have to pay the price that the result depends on the arbitrary matching scale. Although the $t_{\rm ini}$ dependence is canceled at next-to-leading logarithmic level, this dependence could still could be a source of uncertainty [\refcite{Mrenna:2003if}].

\section{Matching parton shower and matrix elements at next-to-leading order level}

Matching the parton shower and the NLO computation is a long outstanding problem. Recently some very important progress has been made. One example is the program of Frixione, Nason, and Webber [\refcite{FrixioneWebberI}], which so far
has been applied to cases with massless incoming partons but not to
cases with massless final state partons. The other example is that of 
[\refcite{nloshowersI}], which concerns three-jet observables in electron-positron annihilation.

In the next we review a general algorithm for matching fix order NLO computation and parton shower.

\subsection{NLO cross section}

Before we discuss the matching at NLO level it is useful to have a brief review of the structure of the NLO fix order calculation. According to the dipole method in the general case the cross section is
\begin{equation}
\label{eq:xsecNLO}
\begin{split}
\sigma_{\rm NLO} = {}& 
\int_N d\sigma^B 
+\int_{N+1}\left[d\sigma^R-d\sigma^A\right]\\
&+\int_{N}\left[d\sigma^B\otimes\bm{I}(\epsilon)+d\sigma^V\right]_{\epsilon=0}
+\int_{N}d\sigma^B\otimes\left[\bm{K} + \bm{P}(\mu_{F})\right]\;\;.
\end{split}
\end{equation}
The first term $d\sigma^B$ is the Born contribution. The second term is an $N+1$ parton phase space integral which gives the contribution of the real radiations $d\sigma^R$. Since $d\sigma^R$ is singular in some region of the phase space and this singularities are regularized by the dipole subtraction terms $d\sigma^A$. The dipole contributions are based on the dipole factorization formulas [\refcite{Catani:1996vz}].
 
The next term is the contribution of the virtual graphs $d\sigma^V$ with $N$ outgoing external partons. They have also infrared divergences in $d=4$ dimension in terms of $1/\epsilon$ (where epsilon is the parameter of the dimensional regularization) but these divergences are cancelled by 
$d\sigma^B\otimes\bm{I}$. Furthermore we have some finite $N$-parton contributions $d\sigma^B\otimes\left[\bm{K} + \bm{P}(\mu_{F})\right]$.

%Note, in Eq.~\eqref{eq:xsecNLO} the parton distribution function are included implicitly.

\subsection{Cross section with shower}

Let us calculate the cross section of an infrared safe $N$-jet quantity. Assuming we know $|\sigma_{N})$ and $|\sigma_{N+1})$ the matrix element improved cross section from Eq.~\eqref{eq:showerexpandDelta} is
\begin{equation}
\label{eq:showerNjet}
\begin{split}
\sbra{F_N}D(&t_{\rm f})\sket{\sigma_{\Delta}(t_{\rm f})}
=\int_{t_{2}}^{t_{\rm f}}\!\!dt_{N}\,
\sbra{F_N}D(t_{\rm f})N(t_{\rm f},t_{N})
W_{\Delta}(t_{\rm f}, t_{N}, t_{2})\sket{\sigma_{N}}
\\
&
+ \int_{t_{2}}^{t_{\rm f}}\!\!dt_{N+1}\,
\sbra{F_N}D(t_{\rm f})U(t_{\rm f}, t_{N+1})
W_{\Delta}(t_{\rm f}, t_{N+1}, t_{2})
\sket{\sigma_{N+1}}\;\;,
\end{split}
\end{equation}
where operator $D(t_{\rm f})$ represents the hadronization and function $F_{N}$ defines the $N$-jet observable. Since $F_{N}$ is an infrared safe quantity it is safe to expand this formulae in $\as$ and drop the power corrections which are provided by the hadronization. After the expansion one can show that 
\begin{equation}
\label{eq:expandNLO}
\begin{split}
\sbra{F_N}D(t_{\rm f})\sket{\sigma_\Delta} = {}& 
\int_N d\sigma^B \left(1+ C + \aspi W_\Delta^{(1)}\right)
+
\int_{N+1}\left[d\sigma^R-d\sigma^A\right]\\
&
+{\cal O}(\as^2)+{\cal O}(1\,\mathrm{GeV}/\sqrt{s})\;\;.
\end{split}
\end{equation}
On the right hand side we have two integral. The second term is the $N+1$ parton integral of the real radiation like in Eq.~\eqref{eq:xsecNLO}. The first term consists of three contributions. We got the Born term $d\sigma^B$. There are some NLO contribution $d\sigma^B W_\Delta^{(1)}$ that comes from the expansion of the Sudakov reweighting operator $W_{\Delta}$. 

The contribution $d\sigma^B C$ is the {\em error term} of the leading color approximation. Since the splitting operator  ${\cal H}(t)$ is valid only at leading color level in the soft gluon limit this error term generally is not zero. It is a leading order contribution but it is only subleading color contribution
\begin{equation}
C = C^{(0)} + \aspi C^{(1)} = {\cal O}\!\left(\frac1{N_{c}^{2}}\right)\;\;.
\end{equation}
On the other hand only next-to-leading logarithms contributes to this term. 
One can neglect this term but with minor redefinition of 
Eq.~\eqref{eq:showerNjet} it is possible to manage that this term vanishes [\refcite{Nagy:2005aa}]. In this proceeding we simply neglect this term.

One can immediately see that the expression on the right hand side of 
Eq.~\eqref{eq:expandNLO} is almost the NLO cross section. We got the Born term and the real contributions right but we missed the virtual and some finite NLO contributions. In order to match the shower and the NLO fix order computation we have to modify Eq.~\eqref{eq:showerNjet}. Thus we have
\begin{equation}
\begin{split}
\sbra{F_N}D(&t_{\rm f})\sket{\sigma_{\rm NLO}(t_{\rm f})}
=\int_{t_{2}}^{t_{\rm f}}\!\!dt_{N}\,
\sbra{F_N}D(t_{\rm f})N(t_{\rm f},t_{N})
W_{\Delta}(t_{\rm f}, t_{N}, t_{2})\sket{\sigma_{N}}
\\
&
+ \int_{t_{2}}^{t_{\rm f}}\!\!dt_{N+1}\,
\sbra{F_N}D(t_{\rm f})U(t_{\rm f}, t_{N+1})
W_{\Delta}(t_{\rm f}, t_{N+1}, t_{2})
\sket{\sigma_{N+1}}
\\
&
+ \int_{t_{2}}^{t_{\rm f}}\!\!dt_{N}\,
\sbra{F_N}D(t_{\rm f})U(t_{\rm f},t_{N})
W_{\Delta}(t_{\rm f}, t_{N}, t_{2})\sket{\sigma_{N}^{NLO}}\;\;.
\end{split}
\end{equation}
Here we have an extra term that is basically the shower evolution of the state $|\sigma_{N}^{NLO})$. This state provides all the missing terms for the NLO matching
\begin{equation}
\begin{split}
  \int[&d\{c\}_{\La,\Lb,N}]
   \sbra{\sigma_{N}^{NLO}}\{p,f,c\}_{\La,\Lb,N}\big) = 
-\aspi W_\Delta^{(1)}\left|{\cal M}_{N}\right|^{2}\\
&+\left|{\cal M}_{N}\right|^{2}\otimes\big(\bm{K} + \bm{P}(\mu_{F})\big)
+\left[\left|{\cal M}_{N}\right|^{2}_{1-loop}
+\left|{\cal M}_{N}\right|^{2}\otimes\bm{I}(\epsilon)\right]_{\epsilon=0}\;\;,
\end{split}
\end{equation}
where $|{\cal M}_{N}|^{2}$ and $|{\cal M}_{N}|^{2}_{1-loop}$ are the tree and 1-loop level matrix element squares, respectively and the insertions operators $\bm{I}(\epsilon)$, $\bm{K}$ and $\bm{P}(\mu_{F})$ are defined according to Eq.~\eqref{eq:xsecNLO}.

However this matching formulae is very {\em simple} and general we still have some caveats. Even if the problem with the error term of leading color approximation is solved the matching is exact only for the $2\to2$ like processes. That is because in the shower the emissions are ordered and we use only the leading order splitting kernels. This ordering cut out a certain region of the available phase space which is usually considered in the fix order NLO calculations. On the other hand this error is negligible because in this region the NLO computation is not reliable.

\subsection{Slicing method at NLO level} 

In order to have an simpler Sudakov reweighting procedure one can do the NLO matching with slicing method. This algorithm is completely worked out for processes in $e^{+}e^{-}$ annihilation and the details can be find in Ref.~[\refcite{Nagy:2005aa}].

\section{Conclusions}

In this paper we outlined a new parton shower algorithm based on the Catani-Seymour dipole factorization formulas. We also introduced a new formalism for describing the shower and we find it very powerful. The main advantages of the this new algorithm are the followings: i) It is Lorentz invariant and covariant formalism. ii) The evolution is a transverse momentum ordered algorithm and it is managed that the partons are always onshell, the momentum conservation is maintained and the phase space weight is the exact weight in steps of the evolution. iii) There are no {\em ambiguous technical} parameters. The only parameter is the infrared cutoff scale parameter. iv) Improved soft gluon treatment. Although the soft gluon limit is still considered in large $N_{c}$ limit we could make some improvements by taking into account not only the final-final but also the initial-final and initial-initial state color connections. 

We derived a more general formulae for matching Born level matrix elements and parton shower. Our matching algorithm is free form any arbitrary matching parameter but we show that the CKKW method is a specialization of our matching procedure.

Since the shower evolution is based on the dipole factorization, the matching of the shower and NLO fix order computation (based on the Catani-Seymour subtraction method) is rather straightforward.

\section*{Acknowledgments}

I am greatful to the organizers of the Ringberg workshop for their invi-
tation as well as for providing a pleasant atmosphere during the meeting.
This work was supported in part by the Swiss National Science Foundation 
(SNF) under contract number 200020-109162 and by the Hungarian Scientific Research Fund grant OTKA T-038240.

\end{document}